\documentclass[11pt]{article}
\usepackage{moriond,epsfig}

\def\al{\alpha}

\def\be{\begin{equation}}
\def\ee{\end{equation}}
\def\bea{\begin{eqnarray}}
\def\eea{\end{eqnarray}}

\def \as {\relax\ifmmode\alpha_s\else{$\alpha_s${ }}\fi}

\def \al #1 {\frac {\as({#1})}{\pi} }
\def \ds #1 {\ooalign{$\hfil/\hfil$\crcr$#1$}}

\def \GeV {\mbox{GeV}}

\def\qt{Q_T}

\begin{document}
\begin{flushright}
BNL-HET-02/13 \\
BNL-NT-02/16 \\
RBRC-249 \\
YITP-SB-02-36\\
\end{flushright}
\vspace*{1.9cm}
\title{ELECTROWEAK VECTOR BOSON PRODUCTION IN JOINT RESUMMATION}

\author{\vspace*{-2mm}\underline{A. KULESZA}$^a$, G. STERMAN$^b$ AND W. VOGELSANG$^c$}

\address{$^a$Department of Physics, Brookhaven National Laboratory\\
          Upton, NY 11973, U.S.A.\\
         $^b$C.N.\ Yang Institute for Theoretical Physics, SUNY Stony Brook\\
          Stony Brook, New York 11794 -- 3840, U.S.A.\\
         $^c$RIKEN-BNL Research Center and Nuclear Theory, Brookhaven National Laboratory,\\
          Upton, NY 11973, U.S.A. }

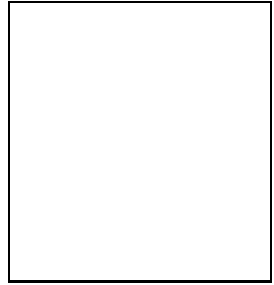
\maketitle\abstracts
{\vspace*{-7mm}We study the application of the joint resummation to electroweak boson
production at hadron colliders. The joint resummation formalism
resums both threshold and transverse momentum corrections to the transverse
momentum distribution at next-to-leading logarithmic accuracy. We obtain a good
description of the transverse momentum distribution of $Z$ bosons produced at
the Tevatron collider.}
\vspace*{-0.5cm}
\vspace*{-4mm}
\section{Introduction}
\vspace*{-2mm}

Electroweak boson production at hadron colliders serves as one of
the major testing grounds for resummation techniques of perturbative QCD. In this
talk we will describe the application of the recently proposed joint
resummation formalism~\cite{LSV} to these type of processes
and compare theoretical 
predictions of the formalism with the data on $Z$ boson production at 
the Tevatron collider.

It is well-known that the partonic subprocesses calculated using perturbation
theory acquire logarithmic corrections due to gluon emission.
The corrections arise from cancellations between virtual and real
contributions at each order in perturbation theory and become large if a
distribution is inspected near the phase-space boundary. 
The two prominent examples are {\em threshold} and {\em recoil} corrections,
encountered in the partonic hard-scattering cross sections. 
The threshold corrections
of the form $\as^n \ln^{2n-1}(1-z)/(1-z)$ become large when the partonic
c.m. energy approaches the invariant
mass $Q$ of the produced boson, $z=Q^2/\hat s \rightarrow 1$. 
The recoil corrections, in turn, are of the form $\as^n
\ln^{2n-1}(Q^2/Q_T^2)$ and grow large if the transverse momentum
carried by the produced boson is very small, $Q_T \ll Q$. Thus,
sufficiently close to the phase-space boundary, i.e. in the limit of soft
and/or 
collinear radiation, fixed-order perturbation theory is bound to fail.
A proper treatment of higher-order corrections in this
limit requires resummation of logarithmic corrections to all orders. Such
techniques are well established both in the threshold~\cite{Sthr,CTthr} 
and in the recoil~\cite{DDT,PP,AEGM,CSS} case.

However, resummation of recoil and threshold corrections separately can
lead to opposite effects, i.e. suppression or enhancement of the partonic cross
section, respectively. A full analysis of soft gluon effects in
transverse momentum distributions $d\sigma/dQ^2\,dQ_T^2$ should therefore
take these two types of corrections simultaneously into account. A joint
treatment of these corrections was proposed in~\cite{LSV,LSVprl}. It relies on a
novel refactorization of short-distance and long-distance physics at fixed
transverse momentum and energy~\cite{LSV}. Similarly to standard threshold and 
recoil resummation, exponentiation of logarithmic corrections occurs in the
impact parameter $b$ space, Fourier-conjugated to transverse momentum $Q_T$
space, and Mellin-$N$ moment space, conjugated to $z$ space. This time 
both transforms are present, resulting in a final expression which
obeys energy and
transverse momentum conservation. Consequently, phenomenological evaluation of the
joint resummation expressions requires providing prescriptions for inverse
transforms from $N$ and $b$ spaces. This also involves  
specifying
a border between resummed perturbation theory and the nonperturbative
regime, by analyzing and parameterizing nonperturbative effects. Moreover, to fully define the expressions a procedure for
matching between the fixed-order and the resummed result needs to be specified.
In the following talk we will discuss these topics in more detail.

\vspace*{-2mm}
\subsubsection{The jointly resummed cross section}
\vspace*{-2mm}

In the  framework of joint resummation~\cite{LSV} we
derive the following expression at next-to-leading logarithmic accuracy 
for electroweak annihilation~\cite{LSV,KSV}:
\bea
\label{crsec}
     \frac{d\sigma_{AB}^{\rm res}}{dQ^2\,dQ_T^2}
     &=&    \sum_a \sigma_{a}^{(0)}\,
\int_{C_N}\, \frac{dN}{2\pi i} \,\tau^{-N}\;    \int \frac{d^2b}{(2\pi )^2} \,
e^{i{\vec{Q}_T}\cdot {\vec{b}}}\, \nonumber \\
&\times&    {\cal C}_{a/A}(Q,b,N,\mu,\mu_F )\;
      \exp\left[ \,E_{a\bar a}^{\rm PT} (N,b,Q,\mu)\,\right] \;
      {\cal C}_{\bar{a}/B}(Q,b,N,\mu,\mu_F) \; ,
\eea
where $\sigma_a^{(0)}$ denotes the Born cross section, $\tau=Q^2/S$, and $Q$
     is the invariant mass of the produced boson.  The flavour-diagonal Sudakov exponent 
$E_{a\bar a}^{\rm PT}$, at the next-to-leading logarithmic (NLL) accuracy in $N$ and
     $b$ was derived in~\cite{KSV}:
\bea
\label{jointsud}
E_{a\bar a}^{\rm PT} (N,b,Q,\mu,\mu_F) &=&
-\int_{Q^2/\chi^2}^{Q^2} {d k_T^2 \over k_T^2} \;
\left[ A_a(\as(k_T))\,
\ln\left( {Q^2 \over k_T^2} \right) + B_a(\as(k_T))\right] \,.
\eea
Dependence on the renormalization scale is implicit in Eq.~(\ref{jointsud})
through the expansion of $\as(k_T)$ in powers of $\as(\mu)$.
$E_{a\bar a}^{\rm PT}$ has the classic form of the Sudakov exponent in the recoil-resummed $\qt$ distribution for
electroweak annihilation, with the $A$ and $B$ functions defined
as perturbative series in $\as$~\cite{DDT,PP,AEGM,CSS}. The coefficients in the 
expansion of these functions are the same as in the pure $\qt$
resummation and are known from comparison with fixed-order calculations~\cite{Kodaira:1982nh,Catani:1988vd,Davies:1984hs};
at the NLL only the logarithmic terms with $A^{(1)},\ B^{(1)}$ and $A^{(2)}$
coefficients contribute,  
\bea
A_a^{(1)} &=&  C_F\; , \qquad\qquad 
B_a^{(1)}\;=\;-\frac{3}{2} C_F\;,\nonumber \\
A_a^{(2)} &=& \frac{C_F}{2} \left[
C_A \left( \frac{67}{18}-\frac{\pi^2}{6} \right) -\frac{10}{9}T_R
N_F\right]\; .
\eea
The quantity $\chi(N,b)$ organizes the logarithms of $N$ and $b$ in 
joint resummation~\cite{KSV}
\be
\label{chinew}
\chi(\bar{N},\bar{b})=\bar{b} + \frac{\bar{N}}{1+\eta\,\bar{b}/
\bar{N}}\; ,
\ee
where $\eta$ is a constant and we define
\bea
\label{nbdefs}
\bar{N} = N{\rm e^{\gamma_E}} \; , \\
\bar{b}\equiv b Q {\rm e^{\gamma_E}}/2 \; ,
\eea
with $\gamma_E$ the Euler constant.
As expected, with this choice of the form for $\chi(\bar{N},\bar{b})$
the LL and NLL terms are correctly reproduced in the threshold limit, 
$N \rightarrow \infty$ (at fixed $b$), and in the recoil limit 
$b \rightarrow \infty$ (at fixed
$N$).

The functions ${\cal C}(Q,b,N,\mu,\mu_F )$ in Eq.~(\ref{crsec})
are chosen to correspond to the jointly resummed cross section for large $N$
and arbitrary $b$, and to  $Q_T$ resummation for $b\to \infty$, $N$ fixed:
\be
\label{cpdf}
{\mathcal C}_{a/H}(Q,b,N,\mu,\mu_F )
=   \sum_{j,k} C_{a/j}\left(N, \alpha_s(\mu) \right)\,
{\cal E}_{jk} \left(N,Q/\chi,\mu_F\right) \,
               f_{k/H}(N ,\mu_F) \; .
\ee
They are products of parton distribution functions $f_{k/H}$ at scale $\mu_F$,
an evolution matrix ${\cal E}_{jk}$ and coefficients
$C_{a/j}(N,\as)$ with a structure of a perturbative series in $\as$.
The first order expansions of the latter are given here by
\bea
C_{q/q}^{(1)}\left( N,\as \right) &=& 1+
\frac{\as}{4\pi} C_F \left(-8+\pi^2 +\frac{2}{N(N+1)} \right) \; =\;
C_{\bar{q}/\bar{q}}^{(1)}\left( N,\as \right)\; ,\\
C_{q/g}^{(1)}\left( N,\as\right) &=&\frac{\as}{2\pi}\frac{1}{(N+1)(N+2)}\; =\;
C_{\bar{q}/g}^{(1)}\left( N,\as\right)\;  .
\eea

The matrix ${\cal E} \left(N,Q/\chi,\mu_F\right)$ represents 
the evolution of the 
parton densities from scale $\mu_F$ to scale $Q/\chi$ up to NLL accuracy~\cite{KSV} in $\ln N$. It is derived from NLO
solutions of standard evolution
equations~\cite{Furmanski:1981cw,Blumlein:1997em}.
By incorporating full evolution of parton densities (as opposed to only
the leading $N$ part of the anomalous dimension) the
cross section~(\ref{crsec}) correctly includes the leading $\as^n
\ln^{2n-1}(\bar N) /N$ collinear non-soft terms to all orders.
In fact, due to our treatment of evolution, expansion of the resummed cross
section~(\ref{crsec}) in the limit $N \rightarrow \infty,\,b=0$ returns all 
${\cal O}(1/N)$ terms in agreement with the fixed-order result.
Further comparison can be undertaken in the limit $b \rightarrow
\infty,\,N=0$ when our joint resummation turns into standard $Q_T$
resummation. Consequently, the NLO transverse momentum distribution is
recovered from the ${\cal O}(\as)$ expansion of the jointly resummed cross
section exactly in the same way as in the $Q_T$ resummation. Outside of these
limits, a numerical 
comparison between the fixed-order 
and the expanded jointly resummed expression for $d \sigma / d Q_T$ at ${\cal
O}(\as)$ shows, as expected, a very good agreement--especially in the small
$Q_T$ region.

\vspace*{-2mm}
\subsubsection{Inverse transforms and matching}
\vspace*{-2mm}

The jointly resummed cross section~(\ref{crsec}) requires defining inverse Mellin and Fourier transforms so that 
singularities associated with the Landau pole are avoided. Similarly to threshold
resummation, there are also singularities in $N$ space associated with parton
distribution functions. A contour for the Mellin integral
in~(\ref{crsec}) is chosen to be analogous to the 'minimal prescription'
contour in threshold resummation~\cite{CMNT} 
\be
\label{cont}
N = C+ z {\rm e}^{\pm i \phi} \; .
\ee
For $\phi>\pi/2$, this results in an exponentially convergent integral
over $N$ in the inverse transform, Eq.~(\ref{crsec})
for all $\tau<1$ \cite{CS,CMNT}. The constant $C$ has to be such that the
contour lies to the right of the rightmost singularity of the parton
distribution functions but left of the Landau pole.

The inverse Fourier integral from $b$ space also suffers from the Landau singularity.
We define this integral using the identity
\be
\label{bint}
\int d^2 b \;e^{i{\vec{Q}_T}\cdot {\vec{b}}}\,f(b)\;=\;  2 \pi\,
\int_0^{\infty} \, db\,b\, \,J_0(bQ_T) \,f(b) \;=\; \pi\, \int_0^\infty db\, b\, \left[\, h_1(bQ_T,v) + h_2(bQ_T,v)\,\right]\,f(b) \, ,
\ee
and employing Cauchy's theorem to deform the integration contour  along the
real $b$ axis into a contour in complex $b$ plane~\cite{LSVprl,KSV}. The auxiliary
functions $h_{1,2}(z,v)$ are related
to Hankel functions and are defined by integrals in the complex 
$\theta$-plane~\cite{grad}:
\bea
h_1(z,v)
&\equiv& - {1\over\pi}\
\int_{-iv\pi}^{-\pi+iv\pi}\, d\theta\, {\rm e}^{-iz\, \sin\theta}\; ,
\nonumber\\
h_2(z,v)
&\equiv& - {1\over\pi}\
\int^{-iv\pi}_{\pi+iv\pi}\, d\theta\, {\rm e}^{-iz\, \sin\theta}\, .
\label{hdefs}
\eea
The $h$ functions distinguish
between the positive and negative phases in Eq.~(\ref{bint}). The $b$
integral can thus be written as a sum of two contour integrals (plus 
contributions vanishing at large $|b|$, coming from closing the contours), one integral
of the corresponding integrand with $h_1$
along a contour $C_1$ in the upper half of the $b$ plane, the other integral
of the integrand with $h_2$ along a 
contour $C_2$ in the lower half.  
The Landau pole can be avoided if one defines the contours in the following way:
\be
\label{upperb}
C_1: \quad b =\left\{ \begin{array}{ll}
   t & (0\leq t\leq b_c) \\
b_c - t  {\rm e}^{-i \phi_b} & (0\leq t\leq \infty)
\end{array} \right. \qquad \qquad
C_2: \quad b =\left\{ \begin{array}{ll}
   t & (0\leq t\leq b_c) \\
   b_c - t  {\rm e}^{i \phi_b} & (0\leq t\leq \infty)
\end{array} \right. \; ,
\ee  
where parameters $b_c$ and $\phi_b$ are arbitrary as long as the contour does
not run into the Landau pole singularity or singularities associated with the
particular form~(\ref{chinew}) of the function $\chi$.

In the joint resummation we adopt the following matching prescription between
the resummed and the fixed-order result:
\begin{equation}
{d \sigma \over d Q^2 d Q_T^2} = {d \sigma^{\rm res} \over d Q^2 d Q_T^2}
-  {d\sigma^{\rm exp(k)} \over d Q^2 d Q_T^2} +
{d \sigma^{\rm fixed(k)} \over d Q^2 d Q_T^2} \,,
\label{joint:match}
\end{equation}
where $d \sigma^{\rm res}/d Q^2 d Q_T^2$ is given in Eq.~(\ref{crsec}) and
$d\sigma^{\rm exp(k)}/d Q^2 d Q_T^2$ denotes the terms resulting from
the expansion of the resummed expression in powers of
$\as(\mu)$ up to the order $k$ at which the fixed-order cross section
$d \sigma^{\rm fixed(k)} /d Q^2 d Q_T^2$  is taken. The above matching prescription
 in $(N,b)$ space guarantees that no double counting of singular contributions
occurs in the matched distribution.

\vspace*{-2mm}
\subsubsection{Transverse momentum distribution for $Z$ production}
\vspace*{-2mm}

\begin{figure}[h]
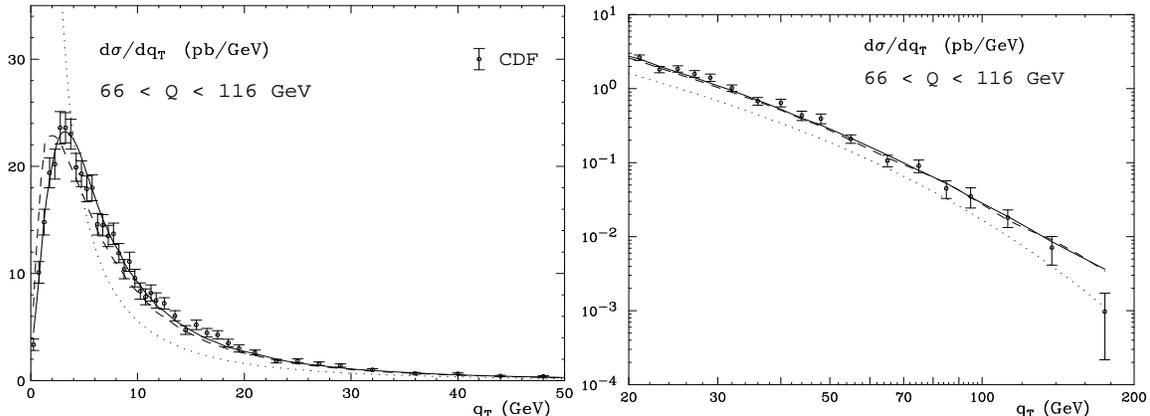

\begin{center}
\includegraphics[width=5.5cm,height=7.5cm,angle=90]{cdf.epsi}
\includegraphics[width=5.5cm,,height=7.5cm,angle=90]{cdf_largeqt.epsi}
\end{center}
\vspace*{-5mm}
\caption{CDF data {\protect \cite{CDF}} on Z production compared to joint resummation
predictions (matched to the ${\cal O}
(\as)$ result according to Eq.~(\ref{joint:match})
without nonperturbative smearing (dashed) and with Gaussian smearing using the
nonperturbative  parameter $g=0.8$ GeV$^2$ (solid).The dotted line shows the fixed-order result. The normalizations
of the curves (factor of 1.035)  have been adjusted in order to give an optimal
description.  We use CTEQ5M parton distribution functions, $\mu=\mu_F=Q$ and
$\phi=\phi_b=25/32 \pi$, $C=1.3$, $b_c=0.2/Q$.}
\label{fig:cdf}
\end{figure}

Joint resummation predictions for $Z$ boson production compared with the
latest CDF data from the Tevatron collider~\cite{CDF} are shown in
Fig.~\ref{fig:cdf}. Due to the contour integral prescription for
performing inverse transforms, in the framework of joint resummation one does
not require any extra nonperturbative information to obtain predictions.
This is not the case in the standard $Q_T$ resummation formalism, where
nonperturbative parameters are introduced to make the theoretical
expression well defined. As shown by the dashed line in Fig.~\ref{fig:cdf} the 
joint resummation  without any extra nonperturbative input already provides
a good description of the data, except for the region of very small $Q_T$, where
the nonperturbative effects are expected to play a significant role.
However, the form of the nonperturbative input can be
predicted within the joint resummation by taking the limit of small
transverse momentum of soft radiation in the exponent, Eq.~(\ref{jointsud}). Assuming
moderate threshold
effects the procedure gives a simple Gaussian
parametrization $F_{NP}(b)=\exp(-g b^2)$. The value of the parameter $g=0.8
\GeV^2$ is determined by
fitting the predicted distribution to the data. It is very similar to the
value obtained in Ref.~\cite{QZ}, where an extrapolation of the $Q_T$-resummed
cross section to large $b$ was made. The solid line in
Fig.~\ref{fig:cdf} represents predictions including the
nonperturbative parametrization.
In the large $Q_T$ region, see Fig.~\ref{fig:cdf}b, the joint resummation formalism with the matching
prescription~(\ref{joint:match}) also returns a very good description of data without
requiring an additional switching to pure fixed-order result, unlike in the
standard $Q_T$ resummation formalism. 

In summary, we obtain a well-defined jointly resummed cross
section, valid for all nonzero $Q_T$, which successfully describes 
the $Q_T$ distribution of $Z$ bosons
produced at the Tevatron.\\

{\small\vspace*{-3mm} {\bf Acknowledgements:} The work of G.S.\ was supported in part by the National
Science Foundation, grants PHY9722101 and PHY0098527. 
W.V.\ is grateful to RIKEN, Brookhaven National Laboratory and the U.S.
Department of Energy (contract number DE-AC02-98CH10886) for support. A.K.\ was supported by the U.S. Department of Energy (contract number DE-AC02-98CH10886).}

\end{document}